\documentclass[prd,superscriptaddress,amsfonts,amssymb,amsmath,showpacs,twocolumn]{revtex4-2}
\usepackage{bm}
\usepackage{amsfonts}
\usepackage{latexsym}
\usepackage[latin1]{inputenc}
\usepackage{graphicx}
\usepackage{amsmath}
\usepackage{palatino}
\usepackage{mathpazo}
\usepackage{textcomp}
\usepackage{float}
\usepackage{pifont}
\newcommand{\cmark}{\ding{51}}%
\newcommand{\xmark}{\ding{55}}%
\usepackage{booktabs}
\usepackage{dcolumn}
\usepackage{ragged2e}
\usepackage{hyperref}
\hypersetup{colorlinks,citecolor=blue}
\hypersetup{colorlinks=true,linkcolor=red,filecolor=magenta,    urlcolor=blue}
\usepackage{amsmath}
\usepackage{xcolor}
\usepackage{orcidlink}
\usepackage[caption=false]{subfig}
\usepackage{commath}
\def\jnl@style{}
\def\aaref@jnl#1{{\jnl@style#1}}

\def\aaref@jnl#1{{\jnl@style#1}}

\def\aj{\aaref@jnl{AJ}}                   % Astronomical Journal
\def\apj{\aaref@jnl{ApJ}}                 % Astrophysical Journal
\def\apjl{\aaref@jnl{ApJ}}                % Astrophysical Journal, Letters
\def\apjs{\aaref@jnl{ApJS}}               % Astrophysical Journal, Supplement
\def\apss{\aaref@jnl{Ap\&SS}}             % Astrophysics and Space Science
\def\aap{\aaref@jnl{A\&A}}                % Astronomy and Astrophysics
\def\aapr{\aaref@jnl{A\&A~Rev.}}          % Astronomy and Astrophysics Reviews
\def\aaps{\aaref@jnl{A\&AS}}              % Astronomy and Astrophysics, Supplement
\def\mnras{\aaref@jnl{Mon.~Not.~Roy.~Astron.~Soc.}}             % Monthly Notices of the RAS
\def\prd{\aaref@jnl{Phys.~Rev.~D}}        % Physical Review D
\def\plb{\aaref@jnl{Phys.~Lett.~B}}        % Physics Letters B
\def\prc{\aaref@jnl{Phys.~Rev.~C}}  % Physical Review C
\def\prl{\aaref@jnl{Phys.~Rev.~Lett.}}    % Physical Review Letters
\def\qjras{\aaref@jnl{QJRAS}}             % Quarterly Journal of the RAS
\def\skytel{\aaref@jnl{S\&T}}             % Sky and Telescope
\def\ssr{\aaref@jnl{Space~Sci.~Rev.}}     % Space Science Reviews
\def\zap{\aaref@jnl{ZAp}}                 % Zeitschrift fuer Astrophysik
\def\nat{\aaref@jnl{Nature}}              % Nature
\def\aplett{\aaref@jnl{Astrophys.~Lett.}} % Astrophysics Letters
\def\apspr{\aaref@jnl{Astrophys.~Space~Phys.~Res.}} % Astrophysics Space Physics Research
\def\physrep{\aaref@jnl{Phys.~Rep.}}      % Physics Reports
\def\physscr{\aaref@jnl{Phys.~Scr}}       % Physica Scripta
\def\commat{\aaref@jnl{Comm.~Math.~Phys.}}              % Communications in Mathematical Physics
\def\science{\aaref@jnl{Science}}               % Science
\def\cqg{\aaref@jnl{Classical Quant.~Grav.}}            % Classical and Quantum Gravity
\def\jpcs{\aaref@jnl{JPCS}}                                     % Journal of Physics Conference Series
\def\ijmpd{\aaref@jnl{Int.~J.~Mod.~Phys.~D}}                    % International Journal of Modern Physics D
\def\grg{\aaref@jnl{Gen.~Relat.~Gravit.}}               % General Relativity and Gravitation
\def\rpp{\aaref@jnl{Rep.~Prog.~Phys.}}          % Reports on Progress in Physics
\def\npa{\aaref@jnl{Nucl.~Phys.~A}}        % Nuclear Physics A
\def\lrr{\aaref@jnl{Living Rev.~Rel.}}                   % Living reviews in relativity
\def\jcap{\aaref@jnl{J.~Cosmology Astropart.~Phys.}}    % Journal of cosmology and astroparticle physics
\def\rmp{\aaref@jnl{Rev.~Mod.~Phys.}}   %Reviews of modern physics
\def\epjc{\aaref@jnl{Eur.~Phys.~J.~C}}

\newcommand{\mao}{Main Astronomical Observatory of the National Academy of Sciences of Ukraine, \\
27 Akademik Zabolotny St., Kyiv, 03143, Ukraine}
\newcommand{\kao}{Astronomical Observatory, Taras Shevchenko National University of Kyiv, \\
3 Observatorna St., 04053 Kyiv, Ukraine}

\allowdisplaybreaks[1]
\begin{document}
%\color{red}
\color{black}       %% For one column

\title{$H_0$ Tension in Torsion-based Modified Gravity}
%\end{document}

\author{Sanjay Mandal\orcidlink{0000-0003-2570-2335}}
\email{sanjaymandal960@gmail.com}
\affiliation{Faculty of Mathematics \& Computer Science, Transilvania University of Brasov, Eroilor 29, Brasov, Romania}
\author{Oleksii Sokoliuk\orcidlink{0000-0003-4503-7272}}
\email{oleksii.sokoliuk@mao.kiev.ua}
\affiliation{\mao}
\affiliation{\kao}
\author{Sai Swagat Mishra\orcidlink{0000-0003-0580-0798}}
\email{saiswagat009@gmail.com}
\affiliation{Department of Mathematics, Birla Institute of Technology and
Science-Pilani,\\ Hyderabad Campus, Hyderabad-500078, India.}
\author{P.K. Sahoo\orcidlink{0000-0003-2130-8832}}
\email{pksahoo@hyderabad.bits-pilani.ac.in}
\affiliation{Department of Mathematics, Birla Institute of Technology and
Science-Pilani,\\ Hyderabad Campus, Hyderabad-500078, India.}
\affiliation{Faculty of Mathematics \& Computer Science, Transilvania University of Brasov, Eroilor 29, Brasov, Romania}
%
%%%%%%%%%%%%%%%%%%%%%%%%%%%%%%%%%%%%%  DATE  %%%%%%%%%%%%%%%%%%%%%%%%%%%%%%%%%%%%
\date{\today}
\begin{abstract}
The rising concern in the Hubble constant tension ($H_0$ tension) of the cosmological models motivates the scientific community to search for alternative cosmological scenarios that could resolve the $H_0$ tension. In this regard, we aim to work on a torsion-based modified theory of gravity which is an alternative description to the coherence model. We solve numerically for the Hubble parameter using two exponential Lagrangian functions of torsion $T$ and a trace of energy-momentum tensor $\mathcal{T}$ for the dust case. Further, we constrain the cosmological and model parameters; to do that, we use Hubble, SNe Ia, Baryon Acoustic Oscillations,  Cosmic Microwave Background samples, and Markov Chain Monte Carlo (MCMC) simulation through Bayesian statistics. We obtain the values of Hubble constant $H_0$ for our model, and the outputs align with the recent observational measurements of $H_0$. In addition, we check the deviation of our results from model-independent measurements of $H_0$ from Planck2018, S$H_0$ES, and $H_0$LiCOW experiments. In contrast, our finding partially solved the $H_0$ tension but gave a new possible direction to alleviate the $H_0$ tension.\\

\end{abstract}

\maketitle

\date{\today}

%%%%%%%%%%%%%%%%%%%%%%%%%%%%%%%%%%%%%%%%%%%%%%%%%%%%%%%%%%%%%%%%%%%%%%%%
%%%%%%%%%%%%%%%        Introduction        %%%%%%%%%%%%%%%%%%%%%%%%%%%%%
%%%%%%%%%%%%%%%%%%%%%%%%%%%%%%%%%%%%%%%%%%%%%%%%%%%%%%%%%%%%%%%%%%%%%%%%
\section{Introduction}

In modern cosmology, the Hubble constant ($H_0$) tension and the nature of the dark energy are two important topic for discussion. Studying and understanding these problems may bring new physics, which motivates scientists in the present time to look into it. So far, various experiments conducted and ongoing to understand the properties of dark energy/accelerated expansion of the universe and it has been more than two decades that cosmologists are trying to find an appropriate model that can assimilate late cosmic acceleration \cite{{Riess/1998},{Perlmutter/1999},{Spergel/2003},{Peebles/2003},{Hinshaw/2013}}. The quest led them to extensive research on gravitational theories. The dark energy model is one of the well-known models, which attempts to explain the late-time acceleration. Einstein's general relativity (GR) is the best choice as a theory of gravity to deal with acceleration of the universe with an additional constant. Still, in GR we often encounter the presence of singularities and it is seen that it degenerates at cosmological distance. The shortcomings of GR led cosmologists to find several modified theories of gravitation such as $f(R)$ gravity, $f(T)$ gravity, modified Gauss-Bonnet gravity $f(G)$, a general coupling between the Ricci scalar and the Gauss-Bonnet $f(R, G)$ gravity(See extensive reviews on cosmological applications in Refs. \cite{{Nojiri/2011},{Felice/2010}}), a coupling between matter and curvature through $f(R,\mathcal{T})$ gravity (where $\mathcal{T}$ is the trace of energy-momentum tensor) \cite{{Harko/2011},{Chakraborty/2013},{Sun/2016},{Singh/2016},{Fayaz/2016}}.

Generally, in modified gravitational theories one generalizes the Einstein-Hilbert action of General Relativity using the curvature description of gravity. However, recently researchers prefer an alternative theory of gravity that uses torsion instead of curvature, called teleparallel gravity \cite{Andrade/2000, Aldrovandi/2012}. This uses a curvature-free  connection, known as Weitzenb$\ddot{o}$ck connection, instead of the Levi-Civit\'{a} connection of GR and vierbein fields instead of a metric field. Einstein introduced the torsion formalism which is equivalent to that of GR, called teleparallel equivalent general relativity (TEGR) \cite{Unzicker/2005, Moller/1961, Pellegrini/1963, Hayashi/1979, Arcos/2004, Maluf/2013}. GR was generalized to $f(R)$ gravity while teleparallel gravity was generalized to the torsion-based $f(T)$ gravity \cite{{Ferraro/2007}, {Linder/2010}}. Although TEGR is equivalent to General Relativity in terms of describing gravity, $f(T)$ is different than $f(R)$ gravity because they form different gravitational modifications. Also, the field equation in $f(T)$ gravity is of the second order which is an advantage over $f(R)$.  Moreover, $f(T)$ gravity has been explored in many interesting areas such as thermodynamics \cite{Salako/2013}, late-time acceleration \cite{Linder/2010}, reconstruction \cite{Bamba/2012}, static solutions \cite{Hamani/2011, Tamanini/2012, Boehmer/2012}, etc. The feasibility of $f(T)$ gravity at the solar system scales has also been investigated, especially by considering deviations from the linear action of TEGR \cite{Iorio/2012}.

Similar to the coupling between matter and curvature through $f(R, \mathcal{T})$ gravity  (where $\mathcal{T}$ is the trace of energy-momentum tensor), $f(T)$ gravity can be generalized into $f(T,\mathcal{T})$ gravity \cite{{Harko/2014},{Arora/2022}}, which is different from all other existing torsion or curvature-based models. The $f(T,\mathcal{T})$ gravity yields an interesting cosmological phenomenon as it describes the expansion history with an initial inflationary phase, a subsequent non-accelerated matter-dominated expansion, and finally late-time accelerating phase \cite{Momeni/2014}. Also, it has been explored in the context of reconstruction and stability \cite{Junior/2016}, growth factor of sub-horizon modes \cite{Farrugia/2016}, quark stars \cite{Pace/2017}.

Moreover, scientists are successfully able to observe the accelerated expansion of the universe, but still, it is unclear about the dark energy. After many studies and observations, we came up with a few basic properties \cite{vale/2021}: 
\begin{itemize}
    \item dark energy acts as a cosmological fluid with the equation of state $\omega\simeq -1$,
    \item dark energy can hardly cluster, unlike dark matter, and it is filtered homogeneously on cosmic scales of the universe.
\end{itemize} 
Interestingly, the $H_0$ tension is closely related to the nature of dark energy and it states that the globally derived $H_0$ value for $\Lambda$CDM model using CMB measurements \cite{1} is $5\sigma$ lower than the Hubble Space Telescope (HST) measurements for the present scenario of the universe \cite{2}. In literature, a large number of studies have been done to solve or relieve $H_0$ tension \cite{3,4,5}. Mostly, the studies on these problems are done based on the $\Lambda$CDM model, whereas these issues are not widely examined through the modified gravity approach. Also, the modified theories of gravitation are well-known for their successful presentation of accelerated expansion of the universe without having any dark energy or cosmological constant problem. Therefore, in this work, we attempt to explore the $H_0$ tension in the framework of torsion-based modified gravity.

This article is presented as follows: we start by introducing the basic formalism of the torsion-based gravity and solve the motion equations for the solution of the Hubble parameter in section \ref{sec2}. After that, we discuss various observational datasets and the methodology, which are used to do the statistical analysis in section \ref{sec3}. The numerical outputs from our analysis are discussed and summarized in section \ref{sec4}. In last, gathering all the outputs, we conclude in section \ref{sec6}.

\section{Basic Equations of $f(T,\mathcal{T})$ gravity}\label{sec2}

We start with the required connection to obtain a torsion-based curvature, which is called Weitzenb$\ddot{o}$ck connection defined as $\tilde{\Gamma}^{\lambda}_{\,\, \nu \mu} \equiv e_{A}^{\,\,\,\lambda}\,{\partial_{\mu}} \,e^{A}_{\,\,\,\nu}$ which leads to zero curvature, unlike the Levi-Civit\'{a} connection which leads to zero torsion. Here $e_{A}^{\,\,\,\lambda}$ and $e^{A}_{\,\,\,\nu}$ are vierbeins. The metric tensor related to these vierbeins is $g_{\mu \nu}(x)=\eta_{AB}\,e^A_{\,\,\mu}(x) \, e^B_{\,\,\nu}(x)$, here the Minkowski metric tensor $\eta_{AB}=diag(1,-1,-1,-1)$.

The Torsion tensor can be defined as,
\begin{equation}
\label{1}
    T^\lambda_{\,\,\mu \nu}=\tilde{{\Gamma}}^\lambda_{\,\,\nu \mu}-\tilde{{\Gamma}}^\lambda_{\,\,\mu \nu}=e_{A}^{\,\,\,\lambda}(\partial_{\mu}e^A_{\,\,\,\nu}-\partial_{\nu}e^A_{\,\,\,\mu}).
\end{equation}
The contorsion tensor $K^{\mu \nu}_{\,\,\,i} \equiv -\frac{1}{2}(T^{\mu \nu}_{\,\,\,i}-T^{\nu \mu}_{\,\,\,i}-T_{i}^{\,\,\,\mu \nu})$ which expresses the difference between Weitzenb$\ddot{o}$ck and Levi-Civit\'{a} connection. Further, we introduce the superpotential tensor $S_{i}^{\,\,\,\mu \nu}$,
\begin{equation}
\label{2}
    S_{i}^{\,\,\,\mu \nu} \equiv \frac{1}{2}(K^{\mu \nu}_{\,\,\, i}+\delta_i^\mu T^{\alpha \nu}_{\,\,\,\alpha}-\delta_i^\nu T^{\alpha \mu}_{\,\,\,\alpha}).
\end{equation}
Using \eqref{1} and \eqref{2} we can obtain the torsion scalar $T$,
\begin{equation}
\label{3}
    T\equiv  {S_{i}}^{\mu \nu} T^i_{\,\,\mu \nu}
      = \frac{1}{4}{T^{i \mu \nu}T_{i \mu \nu}}+ \frac{1}{2}{T^{i \mu \nu}T_{\nu \mu i}}-T_{i \mu}^{\,\,\,i} T^{\nu \mu}_{\,\,\,\nu}.
\end{equation}
The gravitational action for teleparallel gravity can be defined as,
\begin{equation}
\label{4}
    S=\frac{1}{16\pi G}{\int{{d^4}xeT+\int{{d^4}xe\mathcal{L}_m}}}
\end{equation}
where $e=det(e_\mu^{\,\,A})=\sqrt{-g}$, $G$ is the Newton's constant and $\mathcal{L}_m$ is the matter Lagrangian. From TEGR, one can extend the torsion scalar $T$ to $T+f(T)$, resulting in $f(T)$ gravity. Moreover, the function can be extended to a general function of both torsion scalar $T$ and trace of energy-momentum tensor $\mathcal{T}$ which leads to
\begin{equation}
\label{5}
  S=\frac{1}{16\pi G}{\int{{d^4}xe[T+f(T,\mathcal{T})]+\int{{d^4}xe\mathcal{L}_{m}}}},
\end{equation}
where $f(T,\mathcal{T})$ is the extended general function. The above equation represents the gravitational action for $f(T,\mathcal{T})$ gravity.\\\\
Varying the action, given by eq. \eqref{5}, with respect to the vierbeins yields the field equations
%\begin{widetext}
\begin{multline}
 \label{6}
   (1+f_T)\left[e^{-1}\partial_\mu(e\,e_A^{\,\,\,\sigma }S_\sigma^{\,\,i\mu})-e_A^{\,\,\,\sigma} T^{\mu}_{\,\,\nu \sigma} S_{\mu}^{\,\,\nu i} \right]+\\
   \left(f_{TT}\, \partial_{\mu} T+ f_{T \mathcal{T}}\,\partial_{\mu} \mathcal{T}\right)e\,e_{A}^{\,\,\,\sigma} S_{\sigma}^{\,\,i \mu}+e_A^{\,\,\,i}\left(\frac{f+T}{4}\right)\\
-\frac{f_{\mathcal{T}}}{2} \left(e_A^{\,\,\,\sigma} \stackrel{em}{T}_\sigma^{\,\,\,i} +p\, e_A^{\,\,\,i} \right)= 4\pi G\, e_A^{\,\,\,\sigma} \stackrel{em}{T}_\sigma^{\,\,\,i}
\end{multline}
%\end{widetext}
 where $\stackrel{em}{T}_\sigma^{\,\,\,i}$ is the usual energy-momentum tensor, $f_T={\partial f}/{\partial T}$, $f_{T\mathcal{T}}={\partial^2{f}}/{\partial T \partial \mathcal{T}}$.
 
In order to discuss the geometrical structure of the universe, we consider a spatially flat Friedmann-Lemaitre-Robertson-Walker (FLRW) metric,
   \begin{equation}
   \label{7}
       ds^2=dt^2-a^2(t)\delta_{ij}dx^i dx^j,
   \end{equation}
   where $a(t)$ is the scale factor in terms of time. For the above metric, the vierbein field read,
   \begin{equation}
   \label{8}
       e_\mu^{\,\,\,A}=diag(1,a,a,a),
   \end{equation}
and $T=-6 H^2$. For the cosmological fluid distribution, we consider a perfect fluid and it can be written as
\begin{equation}
\label{9}
   \stackrel{em}{T}_\sigma^{\,\,\,i}=diag (\rho_m,\, -p_m,\,-p_m,\,-p_m) ,
\end{equation}
where $p_m$ and $\rho_m$ are pressure and energy density, respectively and $\mathcal{T}$ reads $\rho_m-3p_m$.
Using the above FLRW metric in the field eq. \eqref{6}, we obtain the modified Friedmann equations:
\begin{equation}\
\label{10}
  {H^2=\frac{8\pi G}{3}{\rho_m}-\frac{1}{6}(f+12H^2f_T)+f_\mathcal{T}(\frac{\rho_m+p_m}{3})},
 \end{equation}
  \begin{multline}\
  \label{11}
  {\dot{H}=-4\pi G(\rho_m+p_m)}-\dot{H}(f_T-12H^2f_{TT})\\-H(\dot{\rho_m}-3{\dot{p_m}})f_{T\mathcal{T}}-f_\mathcal{T}(\frac{\rho_m+p_m}{2}).
  \end{multline}
  Now, one could use the above two field equations to study various cosmological scenarios in the context of $f(T,\mathcal{T})$ gravity. To proceed further in our study, we aim to find the solution for Hubble parameter. But, we have two differential equations with more than two unknown functions. Therefore, we considered the dust universe, for which $p=0$ and the corresponding energy density reads,
\begin{equation}
\label{12}
\rho_m=\frac{\rho_{m0}}{a^3}={\rho_{m0}}(1+z)^3,
\end{equation}
where $a(t)=1/(1+z)$. In dust case, $\mathcal{T}$ reduces to $\rho_m$. Moreover, we need to presume a functional form of Lagrangian $f(T,\mathcal{T})$ to study the cosmological scenario of the universe in the framework of $f(T,\mathcal{T})$ gravity. In this study, we shall explore two types of exponential forms of Lagrangian $f(T,\mathcal{T})$.

\subsection{Exponential Model}
For our first model, we consider the following form of $f(T,\mathcal{T})$ as
\begin{equation}
\label{13}
f(T,\mathcal {T})= T e^{\,\alpha\, \frac{T_0}{T}}+\beta \mathcal{T}.
\end{equation}
Using the above assumption in eq. \eqref{9}, we get

\begin{equation}
\label{14}
E^2-e^{\alpha/E^2}\left(2\alpha-E^2\right)=\left(\frac{2+\beta}{2}\right)\Omega_{m0}a^{-3},
\end{equation}
\\
where $\Omega_{m0}=\frac{\rho_{m0}}{3{H_0}^2}$ is the dimensionless density parameter with present Hubble constant $H_0$. 

In order to reduce the complexity, one can use the present scenario to present a model parameter in terms of other parameters. Therefore, from equation \eqref{14} with $z=0$, one can find

\begin{equation*}
    \alpha= 0.5+ \mathcal{W}\left(\frac{-\beta  \Omega_{m0} -2 \Omega_{m0} +2}{4 \sqrt{e}}\right).
\end{equation*}
Here, $\mathcal{W}$ is the  Lambert function. 

\subsection{Square-root Model}
The second model of our consideration is the well-known sqrt-exponential model, which reads as follows:
\begin{equation}
\label{15}
    f(T,\mathcal{T}) = \alpha T_0(1-e^{-\beta \sqrt{T/T_0}}) + \gamma \mathcal{T}
\end{equation}
Where it is obvious that $T_0$ corresponds to the present day value of the torsion scalar. For such case, there is a corresponding field equation:
\begin{equation}
\label{16}
\begin{gathered}
     (H_0^2(2E^2 - 2\alpha + (2\alpha(1 + \beta E))/\exp(\beta E) \\
     - (1 + z)^3(2 + \gamma)\Omega_{m0}))/2
\end{gathered}
\end{equation}
With the present-day constraint on $\alpha$:
\begin{equation}
\label{17}
    \alpha = -\frac{e^{\beta } (\gamma \Omega_{m0}+2 \Omega_{m0}-2)}{2 \left(-\beta +e^{\beta }-1\right)}
\end{equation}
Now we can proceed to the next section and discuss the data used to constrain our models.
\section{Data and Methodology}\label{sec3}

In this section, we shall discuss the observational data sets and the methodology to estimate the bounds of parameters. For this purpose, we use the Hubble measurements, pantheon SNIa, Baryon Acoustic Oscillations, and  Cosmic Microwave Background samples. To calibrate the data sets, we adopt the Bayesian statistical analysis and use the \textit{emcee} package to Markov chain Monte Carlo (MCMC) simulation. More details about the data sets and statistical analysis are further discussed in the following subsections.

\subsection{Cosmic Chronometer (CC) Dataset}

Various observations have been used to observe the cosmological parameters, such as the cosmic microwave background (CMB) from the Wilkinson Microwave Anisotropy Probe team \cite{{Hinshaw/2013},{Komatsu/2011},{Spergel/2007}} and Planck team \cite{Ade/2016}, baryonic acoustic oscillations (BAO) \cite{Eisenstein/2005}, Type Ia supernovae (SNeIa) \cite{{Riess/1998},{Perlmutter/1999}}. Some of the above models depend on values that require the Hubble parameter to be integrated along the line of sight to explore overall expansion through time. The Hubble parameter $H$ is deeply connected to the history of universe expansion. It is defined as 
$H=\frac{\dot{a}}{a}$, where $a$ represents the cosmic scale factor and $\dot{a}$ as the rate of change about cosmic time. The expansion rate $H(z)$ is obtained as
\begin{equation}
\label{18}
H(z)=-\frac{1}{1+z} \frac{dz}{dt}
\end{equation}
where z is the redshift.

%Two methods are often used to estimate the $H(z)$ value at a certain redshift. One is to extract $H(z)$ from line-of-sight BAO data, while the other uses differential age methods. 
Here we have used $31$ points from the differential age (DA) approach in the redshift range $0.07 < z < 2.42$ and presented in Table \ref{table1}.
\begin{center}
\begin{table*}[!htbp]
 \caption{$H(z)$ datasets consisting of 31 data points}
 \label{table1}
    \begin{tabular}{||c|c|c|c||c|c|c|c||}
    \hline
    $z$ & $H(z)$ & $\sigma_H$ & Ref. & $z$ & $H(z)$ & $\sigma_H$ & Ref. \\[0.5ex]
    \hline \hline
    $0.070$ & $69$ & $19.6$ & \cite{Stern/2010} & $0.4783$ & $80$ & $99$ & \cite{Moresco/2016}\\
    \hline
    $0.90$ & $69$ & $12$ & \cite{Simon/2005} & $0.480$ & $97$ & $62$ & \cite{Stern/2010} \\
    \hline
    $0.120$ & $68.6$ & $26.2$ & \cite{Stern/2010} & $0.593$ & $104$ & $13$ & \cite{Moresco/2012} \\
    \hline
    $0.170$ & $83$ & $8$ & \cite{Simon/2005} & $0.6797$ & $92$ & $8$ & \cite{Moresco/2012}\\
    \hline
    $0.1791$ & $75$ & $4$ & \cite{Moresco/2012} & $0.7812$ & $105$ & $12$ & \cite{Moresco/2012}\\
    \hline
    $0.1993$ & $75$ & $5$ & \cite{Moresco/2012} & $0.8754$ & $125$ & $17$ & \cite{Moresco/2012} \\
    \hline
    $0.200$ & $72.9$ & $29.6$ & \cite{Zhang/2014} & $0.880$ & $90$ & $40$ & \cite{Stern/2010} \\
    \hline
    $0.270$ & $77$ & $14$ & \cite{Simon/2005} & $0.900$ & $117$ & $23$ & \cite{Simon/2005} \\
    \hline
    $0.280$ & $88.8$ & $36.6$ & \cite{Zhang/2014} & $1.037$ & $154$ & $20$ & \cite{Moresco/2012} \\
    \hline
    $0.3519$ & $83$ & $14$ & \cite{Moresco/2012} & $1.300$ & $168$ & $17$ & \cite{Simon/2005} \\
    \hline
    $0.3802$ & $83$ & $13.5$ & \cite{Moresco/2016} & $1.363$ & $160$ & $33.6$ & \cite{Moresco/2015} \\
    \hline
    $0.400$ & $95$ & $17$ & \cite{Simon/2005} & $1.430$ & $177$ & $18$ & \cite{Simon/2005} \\
    \hline
    $0.4004$ & $77$ & $10.2$ & \cite{Moresco/2016} & $1.530$ & $140$ & $14$ & \cite{Simon/2005} \\
    \hline
    $0.4247$ & $87.1$ & $11.2$ & \cite{Moresco/2016} & $1.750$ & $202$ & $40$ & \cite{Simon/2005} \\
    \hline
    $0.4497$ & $92.8$ & $12.9$ & \cite{Moresco/2016} & $1.965$ & $186.5$ & $50.4$ & \cite{Moresco/2015}  \\
    \hline
    $0.470$ & $89$ & $34$ & \cite{Ratsimbazafy/2017} &  &  &  &  \\
    \hline
    \end{tabular}
    \end{table*}
\end{center}

The chi-square function is defined to find the constraint values of the parameters $\alpha, \beta$ $H_0, \Omega_{m0}$.
\begin{equation}
\label{19}
\chi_{CC}^2=\sum_{i=1}^{31}\frac{[H_i^{th}(\theta_s,z_i)-H_i^{obs}(z_i)]^2}{\sigma_{CC}^2(z_i)}
\end{equation}
where $H_i^{obs}$ denotes the observed value, $H_i^{th}$ denotes the Hubble's theoretical value, $\sigma_{z_i}$ denotes the standard error in the observed value and $\theta_s= (\alpha, \beta, \lambda$ $H_0, \Omega_{m0})$ is the cosmological background parameter space. For simplicity, we use $H_0= 100\, h$ In addition, we use the following \textit{prior} to our analysis:
\begin{table}[!htbp]
    \centering
    \caption{Priors for parameter space $ \beta, \gamma$ $H_0, \Omega_{m0}$.}
    \begin{tabular}{c c}
    \hline\hline
    Parameter & prior\\\hline
        $H_0$ & (60,80) \\
        $\Omega_{m0}$ & (0,1)\\
        $\beta_{\rm exp}$ & (0,7)\\
        $\beta_{\rm sqrt}$ & (0.01,150)\\
        $\gamma$ & (0.01,10)\\
        \hline \hline 
    \end{tabular}
    
    \label{table2}
\end{table}
  In our MCMC analysis, we used $100$ walkers and $1000$ steps to find out results.
 % The $1-\sigma$ and $2-\sigma$ CL contour plot is presented in fig. \ref{fig1} and 
  The numerical results are presented in Table \ref{table3}, for CC sample.
  
\subsection{Type Ia Supernovae}

For Type Ia supernovae, we have used Pantheon compilation of 1048 points in the redshift range $0.01<z<2.26$ \cite{{Camlibel/2020},{Scolnic/2018}}, which integrates Super-Nova Legacy Survey (SNLS), Sloan Digital Sky Survey (SDSS),Hubble Space Telescope (HST) survey, Panoramic Survey Telescope and Rapid Response System(Pan-STARRS1). The chi-square function is defined as,
\begin{equation}
\label{20}
\chi^2_{SNa}=\sum_{i,j=1}^{1048}\bigtriangledown\mu_{i}\left(C^{-1}_{SN}\right)_{ij}\bigtriangledown\mu_{j},
\end{equation}
Here $C_{SNa}$ is the covariance matrix \cite{Scolnic/2018}, and
\begin{align*}\label{4c}
\quad \bigtriangledown\mu_{i}=\mu^{th}(z_i,\theta)-\mu_i^{obs}.
\end{align*} 
is the difference between the observed value of distance modulus extracted from the cosmic observations and its theoretical values calculated from the model with given parameter space $\theta$. $\mu_i^{th}$ and $\mu_i^{obs}$ are the theoretical and observed distance modulus respectively. The theoretical distance modulus $\mu_i^{th}$ is defined as $\mu_i^{th}(z)=m-M=5LogD_l(z)$ where $m$ and $M$ are apparent and absolute magnitudes of a standard candle respectively. The luminosity distance $D_l(z)$ defined as,
$D_l(z)=(1+z)\int_{0}^z\frac{dz^\ast}{H(z^\ast)}$. To run MCMC, we used the same \textit{priors}, number of walkers, and steps, which are used in CC sample. The numerical results are presented in Table \ref{table3}, for Pantheon sample.
    \begin{figure*}[!htbp]
    \centering
    \includegraphics[scale=1]{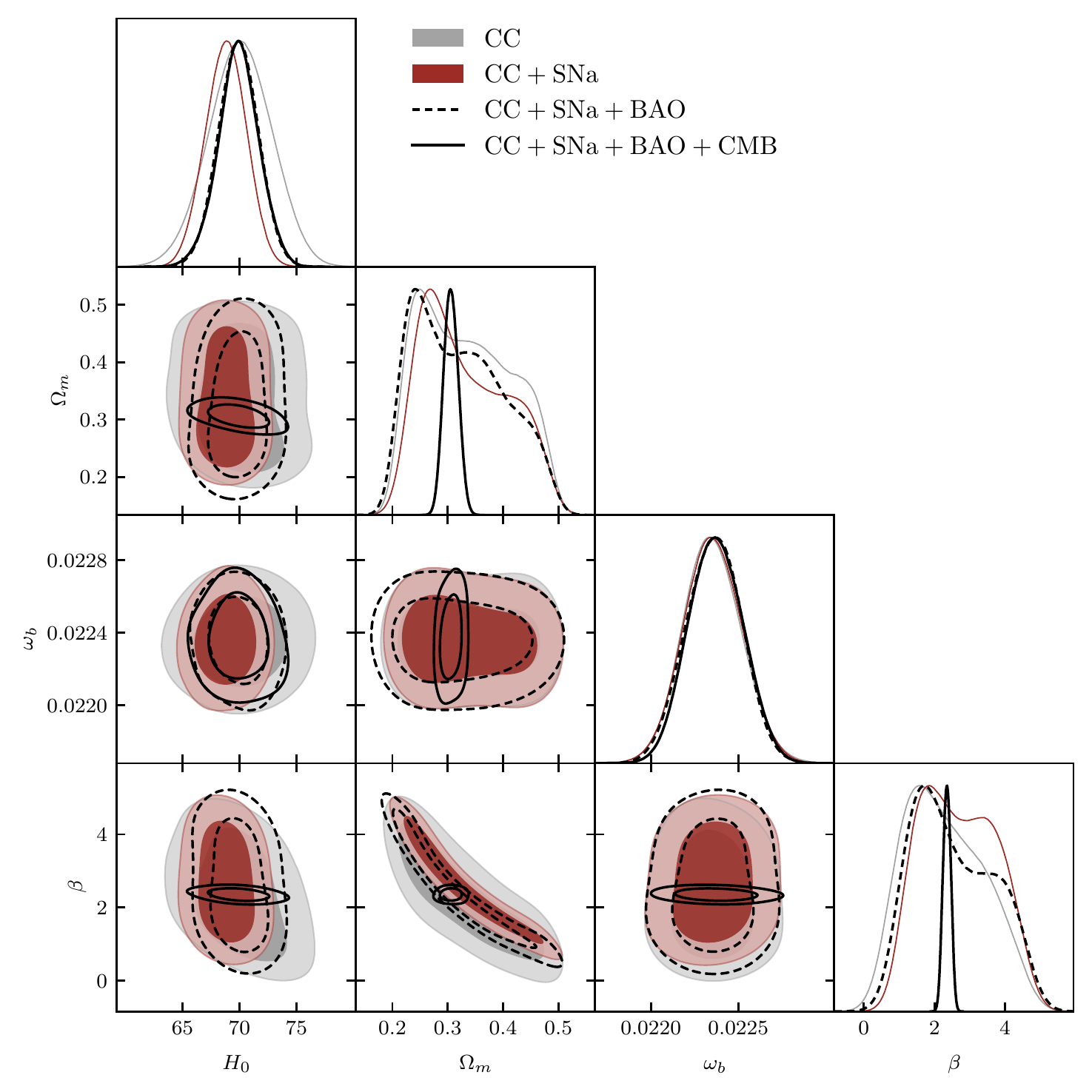}
    
    \caption{The marginalized constraints on the parameters $H_0, \Omega_{m0},\, \beta, \, \omega_b$ of exponential model using various samples. The dark-shaded regions and inner circle  present the $1-\sigma$ confidence level (CL), whereas the light-shaded regions and outer circle present the $2-\sigma$ confidence level.}
    \label{fig1}
\end{figure*}

\begin{figure*}[!htbp]
    \centering
    \includegraphics[scale=1]{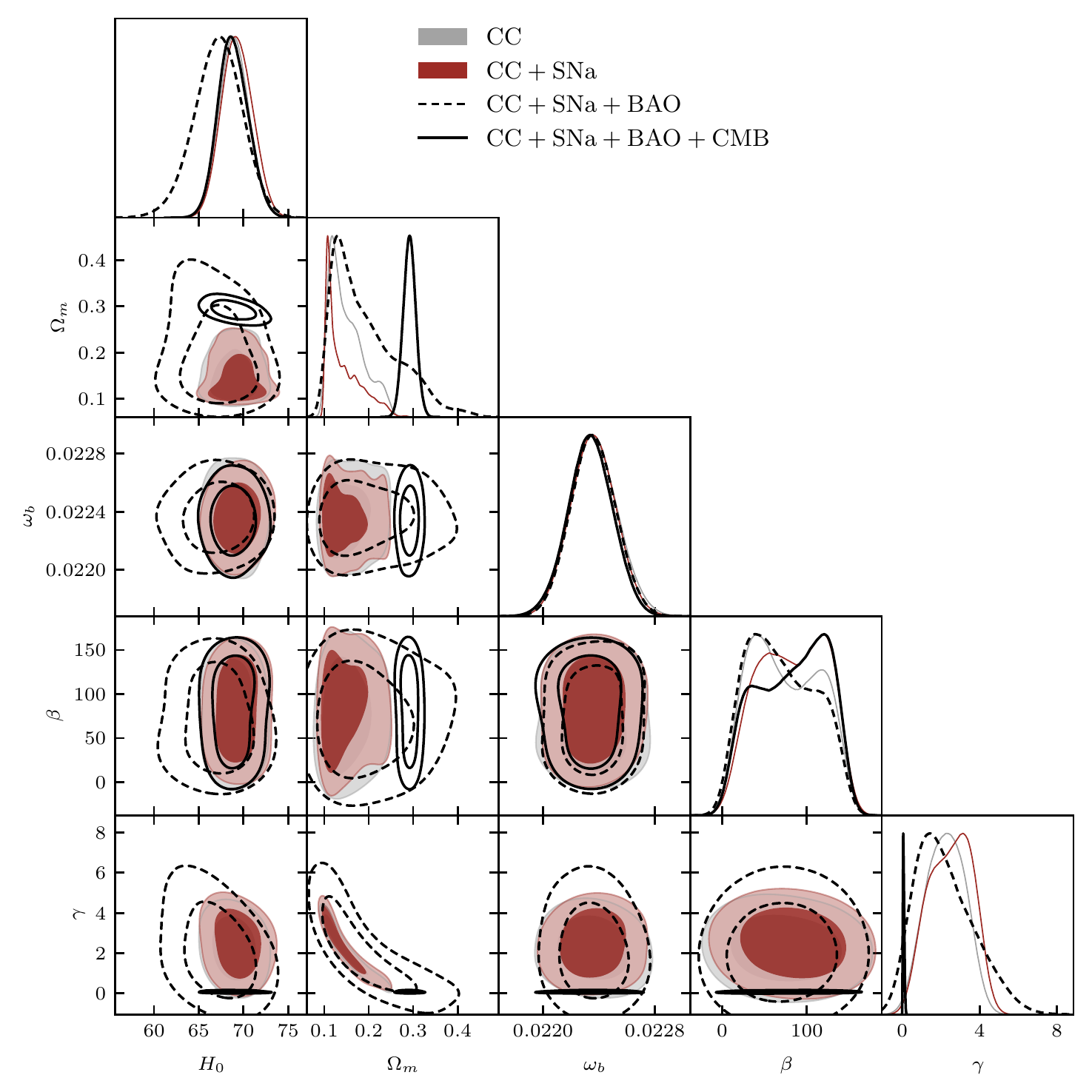}
    
    \caption{The marginalized constraints on the parameters $H_0, \Omega_{m0},\, \beta, \gamma,\, \omega_b$ of square-root model using various samples. The dark-shaded regions and inner circle  present the $1-\sigma$ confidence level (CL), whereas the light-shaded regions and outer circle present the $2-\sigma$ confidence level.}
    \label{fig2}
\end{figure*}

\subsection{Baryon Acoustic Oscillations (BAOs)}

Furthermore, we consider the Baryon Acoustic Oscillations samples to constrain our modified gravity model. BAOs are directly related to the early evolution of the universe, and they can be characterized by the sound horizon $r_s$ at the photon decoupling approach with the redshift $z_*$:
\begin{equation}
\label{21}
    r_s=\frac{c}{\sqrt{3}}\int_0^{\frac{1}{1+z_*}} \frac{da}{a^2H\sqrt{1+(3\Omega_{b0}/4 \Omega_{\gamma 0})a}}
\end{equation}
here, $\Omega_{b0}$ represents the current baryon mass density $z=0$, while $\Omega_{\gamma 0}$ represents the current photon mass density. Furthermore, as previously stated, the angular diameter distance is obtained directly from the BAO sound horizon. For this, we use the $d_A(z_*)/D_V(z_{BAO})$ (here, $d_A(z_*)$ is the angular diameter distance in the comoving coordinates and $D_V(z_{BAO})$ is the dilation scale). The samples for BAO are presented in \cite{BAO}, and we followed the detailed analysis presented in \cite{Simran/2023}.

\subsection{Cosmic Microwave Background}
The last dataset of our consideration is the well-known Cosmic Microwave Background (CMB) observables. We will use compressed data from Planck 2018 results, in which the shift parameters $\mathcal{R}$ and $\ell_a$ are estimated from
\begin{gather}
\mathcal{R}=\sqrt{\Omega_{m0}H_0^2}r(z_\star)/c,\\
\ell_a=\pi r(z_\star)/r_s(z_\star),
\end{gather}
with $r_s$ being the sound horizon and $r$ being the comoving distance to the last scattering surface. We use estimates from Planck 2018 data, as described in \citep{Zhai}, with data vector and covariance data as in Eq. (31) of the work.

\subsection{Joint Analysis}

Lastly, we use different combinations of the above-discussed observational samples. The following combinations we shall use to study.
\begin{align*}
  CC+SNa\\
  CC+SNa+BAO\\
  CC+SNa+BAO+CMB.
\end{align*}
    
The marginalized constraints on the parameters included in the parameter space $\theta$ are presented in fig. \ref{fig1}, \ref{fig2} and numerical results presented in Table \ref{table3}.

\begin{figure}[!htbp]
\centering
    \includegraphics[scale=0.9]{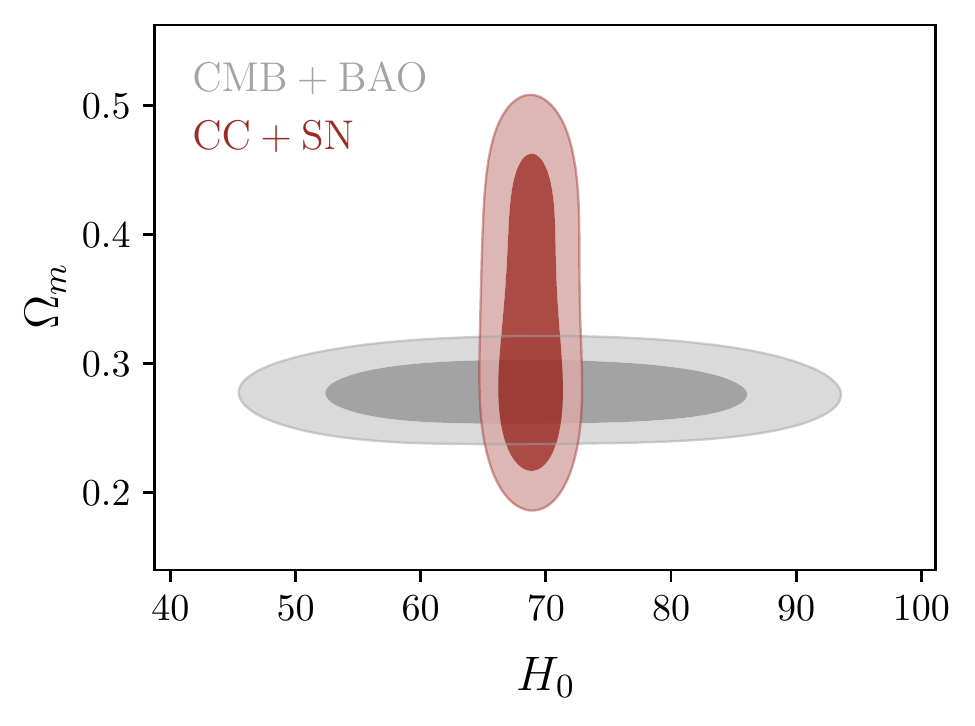}
    \caption{The marginalized constraints on the parameters $H_0, \Omega_{m0}$ of the exponential model using CC+SN and CMB+BAO samples are shown. The dark-shaded regions present the $1-\sigma$ confidence level (CL), and light-shaded regions present the $2-\sigma$ confidence level.}
    \label{first}
\end{figure}

\begin{figure}[!htbp]
\centering
    \includegraphics[scale=0.9]{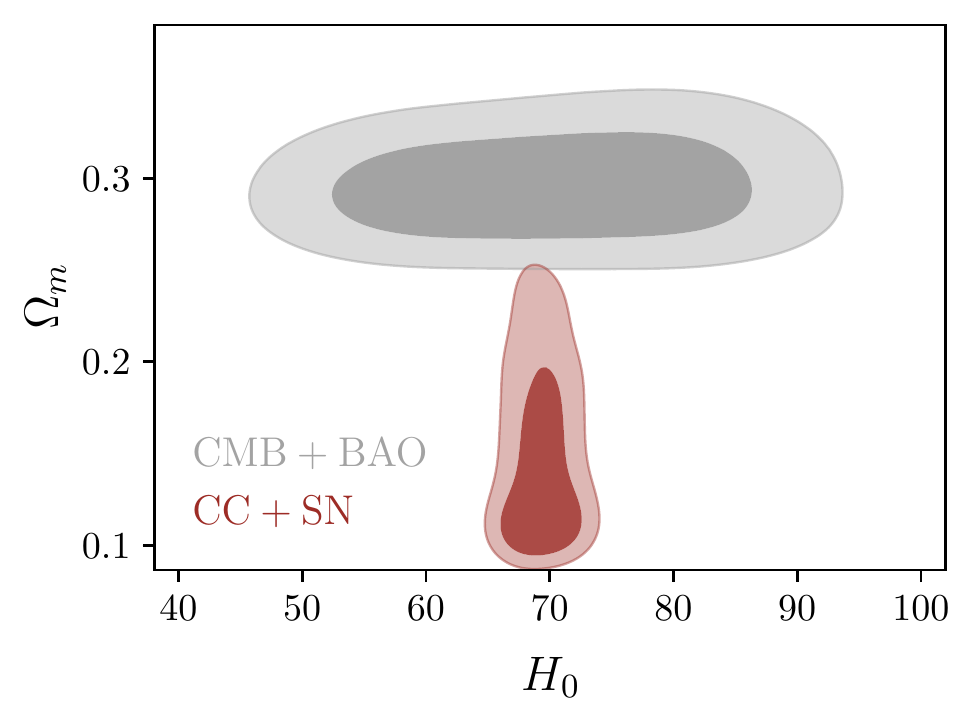}
    \caption{The marginalized constraints on the parameters $H_0, \Omega_{m0}$ of the square-root model using CC+SN and CMB+BAO samples are shown. The dark-shaded regions present the $1-\sigma$ confidence level (CL), and light-shaded regions present the $2-\sigma$ confidence level. }
    \label{second}
\end{figure}

\section{Numerical Results}\label{sec4}

In this section, we shall discuss the numerical results obtained from the statistical analysis. As we know, the $H_0$-tension is a new issue in modern cosmology, because various observational studies presented different values of $H_0$. Therefore, nowadays, the big question is '\textit{why this is happening? or, what is the acceptable range for $H_0$?'}. Let us review the status of the $H_0$ tensions from various experimental outputs. Starting from the 'Gold standard' experimental prediction with the Planck 2018 samples for a flat $\Lambda$CDM model, the Hubble constant is $H_0=67.27\pm 0.60$ km$s^{-1}$ Mp$c^{-1}$ at $68\%$ CL \cite{Aghanim/2020}, and with the addition of the four trispectrum data points to Planck, it is $H_0=67.36\pm 0.54$ km$s^{-1}$ Mp$c^{-1}$ at $68\%$ CL for Planck 2018+ CMB lensing \cite{Aghanim/2020}. The nine-year data released for Wilkinson Microwave Anisotropy Probe (WMAP) experiments \cite{Hinshaw/2013} for the same $\Lambda$CDM model presented a value of Hubble constant $H_0=70.0\pm 2.2$ km$s^{-1}$ Mp$c^{-1}$ at $68\%$ CL. This value is in agreement with previous results for Planck due to the large standard deviation. The above conclusion used to study through various CMP samples, the constraint values of $H_0$ are analyzed for the same $\Lambda$CDM model, such as South Pole Telescope (SPTPol) \cite{Henning/2018} reports $H_0=71.3\pm 2.1$ km$s^{-1}$ Mp$c^{-1}$ at $68\%$ CL  for TE and EE datasets, Atacama Cosmology Telescope (ACT) presents reports 
$H_0=67.9\pm 1.5$ km$s^{-1}$ Mp$c^{-1}$ at $68\%$ CL, ACT with WAMP gives $H_0=67.6\pm 1.1$ km$s^{-1}$ Mp$c^{-1}$ at $68\%$ CL \cite{Aiola/2020}. Finally, a combined analysis of CMB experiments SPT, Atacama Cosmology Telescope Polarimeter (ACTPol), and SPTPol reads $H_0=69.72\pm 1.63$ km$s^{-1}$ Mp$c^{-1}$ at $68\%$ CL \cite{Wang/2020}, while ACTPol+SPTPol+ Planck dataset, gives $H_0=67.49\pm 0.53$ km$s^{-1}$ Mp$c^{-1}$ at $68\%$ CL \cite{Balkenhol/2021}. Moreover, there are some other less precise results presented from the measurements of the polarization of the CMB. These results are $H_0=73.1^{+3.3}_{-3.9}$ km$s^{-1}$ Mp$c^{-1}$ at $68\%$ CL for SPTPol, $H_0=72.4^{+3.9}_{-4.8}$ km$s^{-1}$ Mp$c^{-1}$ at $68\%$ CL for ACTPol, $H_0=70.0\pm 2.7$ km$s^{-1}$ Mp$c^{-1}$ at $68\%$ CL for SPTPol for Planck EE. But, combining these datasets gives $H_0=68.7\pm 1.3$ km$s^{-1}$ Mp$c^{-1}$ at $68\%$ CL \cite{Addison/2021}. Apart from the results obtained from the CMB data analysis, various results were also presented for Baryon Acoustic Oscillations (BAO) and its' combined analysis with CMB considering different cosmological scenarios. For instance, Baryon Spectroscopic Survey (BOSS) Data Release 12 (DR12) provides $H_0=67.9\pm 1.1$ km$s^{-1}$ Mp$c^{-1}$ at $68\%$ CL \cite{iva},  $H_0=68.19\pm 0.36$ km$s^{-1}$ Mp$c^{-1}$ at $68\%$ CL for Planck 2018+ Pantheon Type Ia supernovae+ Dark Energy Survey (DES)+Redshift Space Distortions (RSD)+ Sloan Digital Sky Survey (SDSS) \cite{Alam/2021}, $H_0=68.36^{+0.53}_{-0.52}$ km$s^{-1}$ Mp$c^{-1}$ at $68\%$ CL for WAMP+ BAO \cite{Wang/2020} (please see the articles for more details in $H_0$-tension \cite{{Abbott/2018},{Scolnic/2018},{Troxel/2018},{Krause/2017}}).

\begin{center}
\begin{table*}[!htbp]
\caption{Marginalized constrained data of the parameters $H_0$,\,\,$\Omega_{m0}$, $\beta,\,\,\, \gamma$ for different data samples with 68\% confidence level. }
		    \label{table3}
		\begin{ruledtabular}
		    \centering
		    \begin{tabular}{c c c c c c}
				Model & $H_0$ &$\Omega_{m0}$ & $\Omega_{b0}h^2$& $\beta$ & $\gamma$ \\ 
    	\hline
     & & CC dataset, $68\%$ CL & & & \\
Exponential & $70.08017^{+2.76133}_{-2.69091}$
  & $0.32504^{+ 0.10550}_{- 0.08787}$
 & $0.02235^{+ 0.00017}_{- 0.00016}$
 & $2.13275^{+ 1.44094}_{- 1.12802}$
& - \\
    			Square-root & $67.13177^{+ 2.63125}_{- 2.70886}$
  & $0.16796^{+ 0.10499}_{- 0.05183}$
 & $0.02234^{+ 0.00017}_{- 0.00016}$
 & $69.84053^{+ 55.55565}_{- 45.42065}$
& $1.89607^{+ 1.97057}_{- 1.49719}$
 \\ 
    	\hline
     & & SN dataset, $68\%$ CL & & & \\
				Exponential & $59.87909^{+  6.31530}_{- 5.73575}$
  & $0.21502^{+ 0.05211}_{- 0.03838}$
 & $0.02236^{+ 0.00013}_{- 0.00016}$
 & $4.85598^{+ 1.43277}_{- 1.30637}$
& - \\
    			Square-root & $53.06310^{+ 5.45225}_{- 2.15963}$
  & $0.20367^{+ 0.06537}_{- 0.07138}$
 & $0.02233^{+ 0.00022}_{- 0.00023}$
 & $80.88249^{+ 49.01338}_{- 43.20519}$
& $0.98662^{+ 1.43047}_{- 0.70791}$\\ 
		\hline
     & & CC+SN dataset, $68\%$ CL & & & \\
				Exponential & $68.82544^{+  1.71868}_{- 1.78912}$
  & $0.31303^{+  0.11088}_{- 0.07101}$
 & $0.02236^{+  0.00016}_{- 0.00016}$
 & $2.64907^{+  1.35823}_{- 1.22539}$
& - \\
    			Square-root & $69.19288^{+ 1.89068}_{- 1.71162}$
  & $0.13178^{+ 0.05728}_{- 0.02579}$
 & $0.02235^{+ 0.00016}_{- 0.00015}$
 & $79.44513^{+ 50.30600}_{- 47.81180}$
& $2.48265^{+ 1.22609}_{- 1.36786}$
 \\ 
       \hline
     & & BAO dataset, $68\%$ CL & & & \\
				Exponential & $73.74098^{+ 14.68253}_{- 16.13985}$
  & $0.82048^{+ 0.12978}_{- 0.57898}$
 & $0.02234^{+ 0.00017}_{- 0.00016}$
 & $5.30890^{+ 1.14626}_{- 1.49933}$
& - \\
    			Square-root & $68.05321^{+ 15.51505}_{- 12.62066}$
  & $0.15794^{+ 0.06428}_{- 0.04474}$
 & $0.02239^{+ 0.00015}_{- 0.00018}$
 & $55.28644^{+ 58.71419}_{- 37.21419}$
& $1.80114^{+ 1.57241}_{- 1.13318}$ \\ 
        \hline
     & & CMB dataset, $68\%$ CL & & & \\
				Exponential & $72.87460^{ +11.77759}_{- 14.96356}$
  & $0.56705^{ + 0.28017}_{- 0.29460}$
 & $0.02238^{ + 0.00016}_{- 0.00018}$
 & $2.77895^{ + 0.21634}_{- 0.54019}$
& - \\
    			Square-root & $70.29390^{+ 12.10593}_{- 13.40127}$
  & $0.72886^{+ 0.19418}_{- 0.29493}$
 & $0.02233^{+ 0.00016}_{- 0.00015}$
 & $82.46254^{+ 50.42780}_{- 55.27024}$
& $0.48894^{+ 0.11649}_{- 0.24366}$
 \\ 
    	\hline
     & & BAO+CMB dataset, $68\%$ CL & & &\\
			    Exponential & $69.47906^{+ 13.04077}_{- 13.17503}$
  &$ 0.27687^{+ 0.01794}_{- 0.01782}$
 & $0.02234^{+ 0.00017}_{- 0.00015}$
 & $2.23793^{+ 0.10889}_{- 0.10324}$
& - \\
    			Square-root & $70.52304^{+ 13.12297}_{- 13.77741}$
  & $0.29355^{+ 0.02247}_{- 0.02005}$
 & $0.02239^{+ 0.00015}_{- 0.00017}$
 & $73.04276^{+ 48.61046}_{- 44.18500}$
& $0.07252^{+ 0.05142}_{- 0.03969}$
 \\ 
       \hline
    & & All datasets, $68\%$ CL & & &\\
   				Exponential & $69.96471^{+ 1.70901}_{- 1.61042}$
  & $0.30538^{+ 0.01549}_{- 0.01362}$
 & $0.02236^{+ 0.00016}_{- 0.00017}$
 & $2.35437^{+ 0.12043}_{- 0.10878}$
& - \\
    			Square-root & $68.20605^{+ 1.60699}_{- 3.06440}$
  & $0.29575^{+ 0.01918}_{- 0.01520}$
 & $0.02234^{+ 0.00016}_{- 0.00017}$
 & $118.47522^{+ 18.18268}_{- 37.57121}$
& $0.07850^{+ 0.05231}_{- 0.04218} $ \\
\hline
				Dataset & Model & Planck tension & S$H_0$ES tension& $H_0$LiCOW tension & Viable? \\ 
CC+SN & Exponential & $0.8 \sigma$ & $2.2 \sigma$ & $1.8 \sigma $ & \xmark\\
CC+SN & Square-root & $1.1 \sigma$ & $2.1 \sigma$ & $1.7 \sigma $ & \xmark\\
BAO+CMB & Exponential & $0.2 \sigma$ & $0.3 \sigma$ &$0.3 \sigma$  & \cmark \\
BAO+CMB & Square-root & $0.2 \sigma$ &$0.2 \sigma$ &$0.2 \sigma$  & \cmark \\
All & Exponential & $1.5 \sigma$ &$1.7 \sigma $&$1.4  \sigma $& \cmark \\
All & Square-root &$0.3 \sigma$ &$2.0 \sigma$ &$1.8 \sigma$  & \xmark \\
\end{tabular}
		   \end{ruledtabular}
		\end{table*}
\end{center}
Keeping the issue with Hubble constant $H_0$ in mind, we aim to constrain the $H_0$ in the torsion-based modified theory of gravity framework with CC, Pantheon, BAO, CMB, and their combined samples. In Table \ref{table3}, the numerical outputs for the parameters $ H_0,\, \Omega_{m0},\,\, \beta,\, \gamma$ with $68\%$ CL are presented. From our MCMC analysis, we observed that the Hubble constant as $H_0$ is taking acceptable values in comparison to the recent results for all the samples. Whereas other constraint values of $H_0$ lie in the range of approximately $67$ to $74$, which are aligned with the recent studies on $H_0$ tension. These outputs can be seen in Table \ref{table3}. We also constrain the dimensionless matter density with other parameters of our models. In addition, in the aforementioned table we also add the standard deviations of our results from model-independent measurements of Hubble constant from Planck2018, S$H_0$ES and $H_0$LiCOW experiments. We consider the model to be viable if all tensions are smaller than $2\sigma$. It is worth to notice that under those criteria, only BAO+CMB constraints for both models and Joint constraints for Exponential model satisfy the tensions, mainly due to the relatively large $1\sigma$ bounds on $H_0$ measurement. Those bounds can be reduced in the future studies by using larger amount of late/early universe constraints, such as Big-Bang Nucleosynthesis, Gravitational Wave (mock LISA/LIGO/VIRGO data) and Redshift-Space Distortions data. 
%as $\Omega_{m0}= 0.319^{+0.053}_{-0.072}$ at $68\%$ CL for CC, $\Omega_{m0}= 0.329^{+0.050}_{-0.070}$ at $68\%$ CL for Pantheon, and $\Omega_{m0}= 0.333^{+0.053}_{-0.066}$ at $68\%$ CL for CC+Pantheon dataset. In order to a comparison with a flat $\Lambda$CDM model and to have a better presentation, we constraint the parameters $H_0$ and $\Omega_{m0}$ of $\Lambda$CDM model with CC, Pantheon, and CC+ Pantheon dataset. The numerical outputs are summarized in Table \ref{table3} and $1-\sigma, \, 2-\sigma$ contour presented in fig. \ref{first}, \ref{second}, \ref{third} for respective datasets. Comparing the numerical results of our model with the $\Lambda$CDM model, it is observed that our model's $H_0$ values are slightly higher than the $\Lambda$CDM model.
But, our outputs are in agreement with the observational values of $H_0$ aforementioned and discussed in the review article \cite{vale3}. 

Furthermore, we observed the impact of the model parameter on the value of $H_0$ for each numerical analysis case. It is also observed that the discrepancy in measurements of $H_0$ is large in some cases. At the same time, the $H_0$ reduces for the combined analysis of all datasets. 

  %\end{widetext}

\section{Concluding Remarks}\label{sec6}

The rising concern in the Hubble constant tension ($H_0$ tension) of the $\Lambda$CDM cosmological model motivates the scientific community to search for alternative cosmological scenarios that could resolve the $H_0$ tension. In this view, we have worked on the torsion-based modified theory of gravity to look at this $H_0$ tension issue. For this purpose, we have used the cosmic chronometer dataset, Pantheon Type Ia supernovae samples which integrate various data sets, Baryon Acoustic Oscillations sample, and Cosmic Microwave Background sample. We started by considering two different types of exponential Lagrangian $f(T,\mathcal {T})$ and dust case. Due to complexity, we did the statistical analysis numerically. Further, we constraint the parameters $ H_0,\, \Omega_{m0},\,\, \beta,\, \gamma$ of our model using various observational samples. The Bayesian method is used to find the best-fit ranges of the parameters through MCMC simulation, and constraint values of parameters with $68\%$ CL are presented in Table \ref{table3}.

Moreover, most of the obtained results of $H_0$ have lied in the range of $67$ to $74$ with large discrepancy. Nevertheless, our outputs for $H_0$ are in agreement with discussed $H_0$ values for CMB, BAO experiments in the numerical results section \ref{sec4}. Further, we examined the deviation of our results from model-independent measurements of $H_0$ from Planck2018, S$H_0$ES, and $H_0$LiCOW experiments. Our approach may or may not completely help to resolve the discrepancy in the $H_0$ tension, but it would definitely challenge the theoretician to formulate a $f(T,\mathcal {T})$ Lagrangian in such a way that that would help us to alleviate the $H_0$ tension.

In the concluding note, our findings could motivate the scientific community to look into the $H_0$ tension in the torsion-based gravitational theories as well as other modified theories of gravity. Because our study is one of the alternatives to the coherence model, preferred by the observational dataset, and does not face the cosmological constant problem due to the absence of an additional constant in the presumed Lagrangian $f(T,\mathcal{T})$. In future studies, it would be interesting to see the outputs of these types of studies using weak lensing, LSS spectra, and other datasets. We hope to test and report these types of studies in the near future.\\

\textbf{Data availability:} There are no new data associated with this article.

\section*{Acknowledgements}
 SM acknowledges Transilvania University of Brasov for Transilvania Fellowship for Young Researchers/Postdoctoral research. SSM acknowledges the Council of Scientific and Industrial Research (CSIR), Govt. of India  for awarding Junior Research fellowship (E-Certificate No.: JUN21C05815). PKS  acknowledges the Science and Engineering Research Board, Department of Science and Technology, Government of India for financial support to carry out the Research project No.: CRG/2022/001847 and Transilvania University of Brasov for Transilvania Fellowship for Visiting Professors. We are very much grateful to the honorable referee and to the editor for the illuminating suggestions that have significantly improved our work in terms
of research quality, and presentation.

\end{document}